# Importance of glassy fragility for energy applications of ionic liquids

P. Sippel[1], P. Lunkenheimer[1]*, S. Krohns[1], E. Thoms[1] and A. Loidl[1]

**Ionic liquids (ILs) are salts that are liquid close to room temperature. Their possible applications are numerous, e.g., as solvents for green chemistry[1,2], in various electrochemical devices[3,4] and even for such "exotic" purposes as spinning-liquid mirrors for lunar telescopes[5]. Here we concentrate on their use for new advancements in energy-storage and -conversion devices: Batteries, supercapacitors or fuel cells using ILs as electrolytes could be important building blocks for the sustainable energy supply of tomorrow. Interestingly, ILs show glassy freezing and the universal, but until now only poorly understood dynamic properties of glassy matter[6,7] dominate many of their physical properties[8,9,10,11]. We show that the conductivity of ILs, an essential figure of merit for any electrochemical application, depends in a systematic way not only on their glass temperature but also on the so-called fragility, characterizing the non-canonical super-Arrhenius temperature dependence of their ionic mobility.**

Various new application-relevant material classes have been discovered during recent years. Among them, ILs maybe sparked the largest interest and the number of publications treating these materials by far exceeds those on such prominent topics as high-$T_c$ superconductivity, colossal-magnetoresistance or multiferroicity. For example, as ILs are exclusively composed of ions, they are good candidates for any applications where high ionic conductivity is needed. They also fulfil various other requirements as a broad electrochemical stability window, low vapour pressure or non-flammability[3,4]. The vast number of possible combinations of anions and cations, nowadays known to form ILs, opens up many possibilities for finding compounds optimized for application. Various systematic investigations of the dependence of the conductivity on composition have been performed, e.g., by varying the anion for ILs with the same cation or vice versa[12,13,14,15,16]. However, when considering the fact that there is an estimated number of one million binary ILs[1], obviously these investigations barely have scratched the surface.

Under cooling, ILs usually solidify via a glass transition, a continuous increase of viscosity instead of the abrupt crystallization via a first-order phase transition found in most electrolytes. This also affects their dc conductivity $\sigma_{dc}$ (refs. 8,9,11,13,14,15,16,17,18,19), which exhibits the typical non-canonical temperature dependence known to govern the dynamics of molecules, ions or any other constituents forming glassy matter. In most ionic conductors, the Arrhenius law, $\sigma_{dc} \propto \exp[-E/(k_B T)]$, characteristic for thermally activated ion hopping over an energy barrier $E$, provides a good description of this temperature dependence. However, in ILs it can be parameterized by the empirical Vogel-Fulcher-Tammann (VFT) formula known from glass physics, usually written in the modified form[6]:

$$\sigma_{dc} = \sigma_0 \exp\left[\frac{-D T_{VF}}{T - T_{VF}}\right] \quad (1)$$

Here $D$ is the so-called strength parameter, which is used in the classification scheme for glass formers, introduced by Angell[6], to distinguish between so-called strong and fragile glass formers. While the latter exhibit marked deviations from Arrhenius behaviour (small $D$; solid lines in Fig. 1), strong glass formers (dashed lines) more closely follow the Arrhenius law. Another, nowadays more common way to parameterize these deviations is the fragility index $m$ (ref. 7), defined as the slope at $T_g$ in the Angell plot[6], log $y$ vs. $T_g/T$, where $T_g$ is the glass temperature and $y$ is any quantity coupling to the glassy dynamics (viscosity, relaxation time or, in our case, conductivity). $D$ and $m$ are related by $m = 16 + 590/D$ (ref. 7).

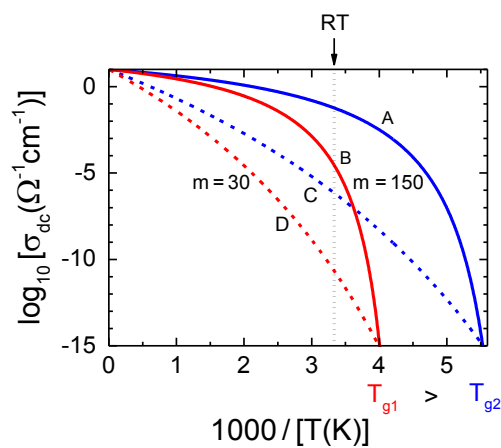

**Figure 1 | Influence of glass temperature and fragility on the room-temperature conductivity of ionic liquids.** The figure shows the temperature dependence of the conductivity (in Arrhenius representation) of four hypothetical ionic glass formers A - D with different glass temperatures and fragilities. Samples A and B, having identical, high fragilities ($m$ = 150 or $D$ = 4.4), demonstrate the effect of different glass temperatures. Glasses A and C (or B and D) having identical glass temperatures but different fragilities (glasses C and D have $m$ = 30 and $D$ = 42) demonstrate the significant influence of the latter quantity on the conductivity. The dotted line indicates room temperature.

It is obvious and well known that the glass temperature $T_g$ has an immediate effect on the conductivity of ILs[20]. It marks the solidification of a glass former when its viscosity reaches about $10^{12}$ Pa s or when the time scale of the molecular motions becomes larger than 100 - 1000 s. Low glass temperatures imply low viscosities at room temperature and in ILs viscosity and conductivity are coupled, even though the degree of coupling can vary[8,11,19,21,22]. Thus the conductivity becomes enhanced for small $T_g$. This is illustrated by curves A and B in Fig. 1, showing temperature-dependent conductivities of four hypothetical ILs. (For many ionic conductors, $\sigma_{dc}(T_g) \approx 10^{-15}$ $\Omega^{-1}$cm$^{-1}$ (ref. 23) and, thus, in Fig. 1 $T_g$ is determined by the point where the curves reach the abscissa). However, based on the phenomenology of glass physics, the fragility should also play an important role (cf. curves A and C in Fig. 1). Indeed, the dependence of the room-temperature viscosity on fragility was already pointed out[8,11]. In the present work we demonstrate that the fragility also has strong impact on the *conductivity* of ILs.

---

[1]Experimental Physics V, Center for Electronic Correlations and Magnetism, University of Augsburg, 86159 Augsburg, Germany
*e-mail: peter.lunkenheimer@physik.uni-augsburg.de

**Table 1 | Parameters of the ionic liquids included in Fig. 4.** Glass temperature ($T_g$), dc resistivity at room temperature ($\rho_{dc}$) and fragility ($m$) are listed. For the materials taken from literature, the reference number is provided. The other liquids were measured in the present work.

| Ionic liquid | $T_g$ (K) | $\rho_{dc}$ ($\Omega$cm) | $m$ | ref. |
|---|---|---|---|---|
| 1-Methyl-3-octylimidazolium hexafluorophosphate (Omim PF$_6$) | 194 | 3.6×10$^3$ | 78 | |
| 1-Hexyl-3-methylimidazolium hexafluorophosphate (Hmim PF$_6$) | 192 | 2.1×10$^3$ | 84 | |
| 1-Butyl-3-methylimidazolium hexafluorophosphate (Bmim PF$_6$) | 189 | 7.7×10$^2$ | 92 | |
| 1-Butyl-3-methylimidazolium bis-(trifluoromethylsulfonyl)imide (Bmim TFSI) | 181 | 2.1×10$^2$ | 88 | |
| 1-Butyl-3-methylimidazolium tetrafluoroborat (Bmim BF$_4$) | 182 | 2.3×10$^2$ | 93 | |
| 1-Butyl-3-methylimidazolium chloride (Bmim Cl) | 228 | 4.2×10$^4$ | 97 | |
| 1-Butyl-3-methylimidazolium tetrachloroferrate (Bmim FeCl$_4$) | 182 | 1.5×10$^2$ | 144 | |
| 1-Butyl-3-methylimidazolium bromotrichloroferrate (Bmim FeCl$_3$Br) | 180 | 1.1×10$^2$ | 146 | |
| 1-Benzyl-3-methyl-imidazolium chlorid (Benzmim Cl) | 253 | 1.4×10$^6$ | 78 | |
| 1-Ethyl-3-methyl-imidazolium tricyanomethanide (Emim TCM) | 183 | 4.6×10$^1$ | 158 | |
| 1-Butylpyridinium tetrafluoroborat (Bpy BF$_4$) | 195 | 4.6×10$^2$ | 117 | |
| (1-Butylpyridinium)$_{0.6}$(1-Butyl-3-methylimidazolium)$_{0.4}$ tetrafluoroborat ([Bpy+Bmim]BF$_4$) | 191 | 3.6×10$^2$ | 111 | |
| 1,3-Dimethylimidazolium(Li 1.0m) bis-(trifluoromethylsulfonyl)imide ([Li+Dimim]TFSI) | 202 | 6.9×10$^2$ | 145 | |
| 1-Propyl-3-methylimidazolium tetrafluoroborate | 175 | 1.9×10$^2$ | 78 | 14 |
| 1-Butyl-3-methylimidazolium tetrafluoroborate | 178 | 2.8×10$^2$ | 93 | 14 |
| 1-Pentyl-3-methylimidazolium tetrafluoroborate | 183 | 6.1×10$^2$ | 78 | 14 |
| 1-Hexyl-3-methylimidazolium tetrafluoroborate | 188 | 8.5×10$^2$ | 66 | 14 |
| 1-Hepyl-3-methylimidazolium tetrafluoroborate | 186 | 1.5×10$^3$ | 68 | 14 |
| 1-Octyl-3-methylimidazolium tetrafluoroborate | 190 | 1.7×10$^3$ | 62 | 14 |
| 1-Nonyl-3-methylimidazolium tetrafluoroborate | 191 | 2.4×10$^3$ | 55 | 14 |
| 1-Hexyl-3-methylimidazolium bis-(trifluoromethylsulfonyl)imide | 187 | 5.9×10$^2$ | 57 | 14 |
| 1-Hexyl-3-methylimidazolium chloride | 220 | 3.7×10$^4$ | 64 | 15 |
| 1-Hexyl-3-methylimidazolium bromide | 216 | 1.1×10$^4$ | 71 | 15 |
| 1-Hexyl-3-methylimidazolium iodide | 208 | 6.3×10$^3$ | 79 | 15 |
| 1-Hexyl-3-methylimidazolium hexafluorophosphate | 194 | 2.4×10$^3$ | 94 | 15 |
| 1-Hexyl-3-methylimidazolium tetrafluoroborate | 187 | 6.9×10$^2$ | 89 | 15 |
| 1-Butyl-3-methylimidazolium bromide | 221 | 1.3×10$^4$ | 69 | 16 |
| 1-Butyl-3-methylimidazolium iodide | 215 | 3.6×10$^3$ | 55 | 16 |
| 1-Butyl-3-methylimidazolium thiocyanate | 195 | 2.4×10$^2$ | 56 | 16 |
| 1-Butyl-3-methylimidazolium tetrafluoroborate | 189 | 3.1×10$^2$ | 56 | 16 |
| 1,3-Dimethylimidazolium dimethylphosphate | 201 | 4.5×10$^2$ | 110 | 17 |
| 1,5-Bis(3-benzyl-2-methylimidazolium)pentane di-bis(trifluoromethanesulfonyl)imide | 250 | 7.1×10$^4$ | 173 | 18 |
| 1,10-Bis(2,3-methylimidazolium)decane di-bis(trifluoromethanesulfonyl)imide | 225 | 4.3×10$^4$ | 141 | 18 |
| 1,10-Bis(3-methylimidazolium)decane di-bis(trifluoromethanesulfonyl)imide | 213 | 1.7×10$^4$ | 141 | 18 |
| 1,5-Bis(3-methyl-2-phenylimidazolium)pentane di-bis(trifluoromethanesulfonyl)imide | 251 | 4.0×10$^4$ | 168 | 18 |

For this purpose, we have measured the dielectric response of 13 ILs. Covering frequencies from about $10^{-1}$ to $10^9$ Hz and a wide temperature range, extending deep into the liquid regime and approaching $T_g$ at low temperatures, enables a thorough analysis of the dc conductivity and the fragility. As a typical example, Figure 2 shows spectra of the real and imaginary part of the dielectric permittivity ($\varepsilon'$ and $\varepsilon''$, respectively), the conductivity ($\sigma'$) and the imaginary part of the dielectric modulus[24] ($M''$) of Omim PF$_6$ (for a definition of the sample abbreviations, see Table 1) for selected temperatures. It should be noted, that partly the information contained in these plots is redundant (e.g., $\sigma' \propto \varepsilon'' \nu$). However, the various dielectrically active processes of ILs are differently emphasized in these plots, making their separate discussion helpful.

In $\varepsilon'(\nu)$ (Fig. 2a) the huge increase and the approach of a plateau at low frequencies is due to electrode polarization or blocking electrodes arising from the trivial fact that the ions cannot penetrate into the metallic capacitor plates[25]. The steep low-frequency increase in the loss, $\varepsilon''(\nu)$, up to about $10^5$ (Fig. 2b), is caused by the dc conductivity $\sigma_{dc}$ and corresponds to the plateau observed in $\sigma'(\nu)$ (Fig. 2c). The levelling off of $\varepsilon''(\nu)$ at the lowest frequencies and the corresponding decrease in $\sigma'(\nu)$ also arises from blocking electrodes. Curiously, in ionic conductors ac measurements are necessary to determine the dc conductivity.

A closer inspection of the rather weakly frequency-dependent regions, observed in $\varepsilon'(\nu)$ and $\varepsilon''(\nu)$ at the higher frequencies, reveals signatures of relaxational processes, i.e. peaks in the loss, accompanied by steps in the real part, both shifting towards higher frequencies with increasing temperatures. This resembles the reorientational relaxations found, e.g., in molecular dipolar glass formers[26]. Such behaviour has also been reported for ILs[9,13,15,27] where it can be ascribed to the reorientational motions of dipolar ions and the corresponding secondary relaxations[9]. In $\sigma'(\nu)$ (Fig. 2c) these modes cause the observed increase at high frequencies, following the dc plateaus. A more detailed assessment of these dipolar modes is out of the scope of the present work.

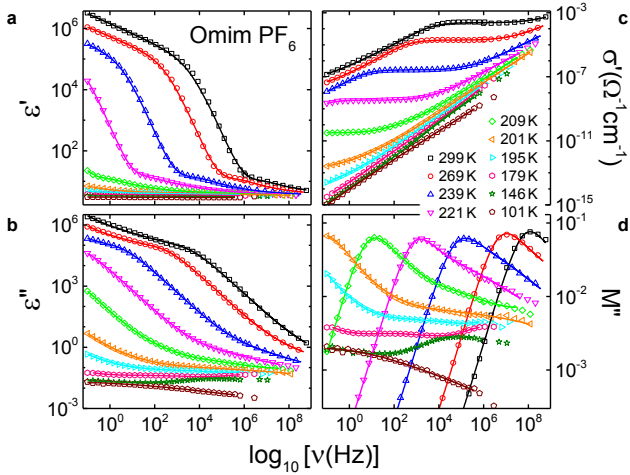

**Figure 2 | Dielectric spectra of Omim PF$_6$.** Spectra are included for a variety of temperatures. The shown quantities are: Dielectric constant (**a**), dielectric loss (**b**), conductivity (**c**) and the imaginary part of the dielectric modulus (**d**). The lines in **a** and **b** are fits assuming a distributed RC circuit to model the blocking electrodes[25], dc conductivity and three relaxational processes described by the Cole-Davidson or Cole-Cole functions. $\varepsilon'(\nu)$ and $\varepsilon''(\nu)$ were simultaneously fitted. The lines in **c** and **d** were calculated from the fits to $\varepsilon'$ and $\varepsilon''$.

$M''$ spectra as shown in Fig. 2d are often used to define a so-called conductivity relaxation time $\tau_\sigma$, thought to characterise the ionic dynamics[24]. The main peak in $M''(\nu)$ is ascribed to the ionic motions and $\tau_\sigma$ is determined from the peak frequency $\nu_\sigma$ via $\tau_\sigma = 1/(2\pi\nu_\sigma)$. The modulus representation suppresses the effects of blocking electrodes showing up at low frequencies in the other quantities. It is known that relaxation peaks observed in $\varepsilon''(\nu)$ also lead to peaks in $M''(\nu)$, but with somewhat shifted peak frequency[28]. Thus the additional peaks and shoulders observed in Fig. 2d at higher frequencies have the same origin as those found in Fig. 2b. One should be aware that the significance of the modulus representation is rather controversial but, nevertheless, often used in ILs[9,13,17,19,21].

We have fitted the spectra of Fig. 2 assuming a distributed RC circuit for the blocking electrodes[25], dc conductivity and three reorientational modes. For the latter, the empirical functions usually applied to molecular glass formers as the Cole-Davidson (for the main reorientational mode) or Cole-Cole function (for the secondary relaxations)[26,29] were used. An excellent description of the experimental data over ten decades of frequency is possible in this way (lines in Fig. 2). All 13 ILs investigated in the present work (see Table 1 for a list) exhibit qualitatively similar behaviour as shown in Fig. 2. In literature, similar spectra are also documented for other ILs[9,13,15,17,20,22]. In the context of the present work, the most interesting quantities determined from these data are $\sigma_{dc}$ and $\tau_\sigma$. The resistivity values $\rho_{dc} = 1/\sigma_{dc}$ obtained by us are listed in Table 1, also including literature data for various additional ILs.

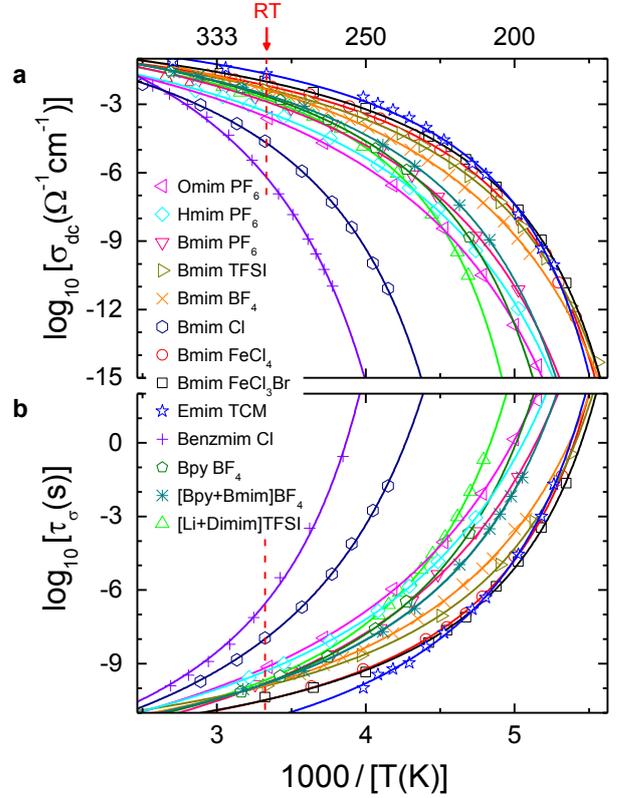

**Figure 3 | Temperature dependence of ionic dynamics.** Data are shown for 13 ILs using an Arrhenius representation. Frame **a** presents the dc conductivity determined from the dielectric spectra. In **b** the conductivity relaxation time deduced from $M''(\nu)$ is provided. The solid lines in **a** and **b** are fits with the VFT formulae, equations (1) and (2), respectively. The dashed lines indicate room temperature. For the meaning of the sample abbreviations in the legend, see Table 1.

Figure 3 shows an Arrhenius representation of the temperature dependences of $\sigma_{dc}$ and $\tau_\sigma$ for the 13 ILs investigated by us. The found non-linear behaviour reveals different degrees of deviations from the Arrhenius law. From $\tau_\sigma$ the glass temperature can be determined using $\tau_\sigma(T_g) = 100$ s (sometimes 250 or 1000 s are assumed, leading to a shift of $T_g$ of few K only). The obtained values are listed in Table 1. It should be noted that this defines the glass temperature of the ionic subsystem. It should agree with the "structural" glass temperature defined, e.g., via the viscosity if charge transport and viscosity are closely coupled. In any case, it is just this glass temperature which determines the conductivity at room temperature (Fig. 1). It reasonably matches the glass temperature read off from the $\sigma_{dc}$ plot of Fig. 3a using the above-mentioned value at $T_g$ of $10^{-15}$ $\Omega^{-1}$cm$^{-1}$ (ref. 23).

The data shown in Fig. 3a were fitted with equation (1) (solid lines). The $\tau_\sigma$ data (Fig. 3b) were fitted with the corresponding VFT formula for the relaxation time:

$$\tau_\sigma = \tau_0 \exp\left[\frac{DT_{VF}}{T - T_{VF}}\right] \qquad (2)$$

At high temperatures, more data points are available for $\sigma_{dc}$ than for $\tau_\sigma$ because the modulus peaks start shifting out of the available frequency window (Fig. 2d) and the peak frequency cannot be unequivocally determined. Therefore we used $D$ as obtained from the fits of $\sigma_{dc}$ to calculate $m$. In some cases the phenomenological VFT fits show significant deviations from the experimental data at low temperatures (e.g., for Bmim TFSI). In such cases we used the original definition of $m$ by the slope in the Angell plot for its determination. The deduced values are provided in Table 1, together with data determined from literature.

Finally, Fig. 4a shows the dependence of the dc resistivity at 300 K on $m$ and $T_g$ for all ILs investigated in the present work and for those from literature. It reveals a clear trend of $\rho_{dc}$ to become strongly reduced at high values of $m$ and low values of $T_g$. The filled black circles are the projections to the two vertical planes of this 3D plot. Obviously, considering the dependence on $T_g$ or $m$ alone, without taking care of the other parameter, is not sufficient to account for the observed resistivity variation. This becomes especially obvious when considering cases where substantial variations of $\rho_{dc}$ show up, despite the corresponding $T_g$ or $m$ values are nearly identical (see, e.g., variation for $m \approx 145$ at the $\rho_{dc}$-$m$ plane or for $T_g \approx 180$ K at the $\rho_{dc}$-$T_g$ plane).

Figure 4b shows the dependence of $\rho_{dc}(300\text{ K})$ on $m$ and $T_g$ as predicted on the basis of the VFT law. To calculate this colour-coded plane, we used eq. (2) with $D$ replaced by $590/(m-16)$ (ref. 7). When assuming $\tau_\sigma(T_g) = 100$ s and $\tau_0 = 10^{-14}$ s, the latter being a typical inverse attempt frequency, one arrives at:

$$\log_{10}[\tau_\sigma(300\text{K})] = \frac{T_g}{300/16 + [300 - T_g]\ln 10 (m-16)/590} - 14 \quad (3)$$

From this the dc resistivity was calculated via $\rho_{dc} = \tau_\sigma/(\varepsilon_\infty \varepsilon_0)$ (ref. 24). Here $\varepsilon_0$ is the permittivity of vacuum and $\varepsilon_\infty$ is the high-frequency limit of the dielectric constant. We used $\varepsilon_\infty = 3$, which is a reasonable average value for ILs. The obtained plane shown in Fig. 4b nicely reproduces the experimentally observed trend in Fig. 4a. The colours shown on the $m$-$T_g$ plane of Fig. 4a correspond to those of the calculated plane plotted in Fig. 4b. Obviously, this calculation can astonishingly well account even for the *absolute* values of the experimentally observed $\rho_{dc}$, despite the various simplifying assumptions made and some uncertainties in the determination of $m$ for data sets that do not extend down to $T_g$.

Notably, the investigated ILs also include one example containing additional lithium ions [1,3-Dimethylimidazolium bis(trifluoromethane)sulfonimide) with 0.1 mol Li]. Such systems are highly relevant for applications in batteries and obviously the correlation of the resistivity with $T_g$ and $m$ also holds for this system. However, more investigations are needed to check the universality of this approach for such ternary systems. Finally, it should be noted that the ILs considered in this work are non-protic. Protic systems are known to show a stronger variation of ionicity[11]. Thus it would be interesting to check if the found correlation holds anyhow.

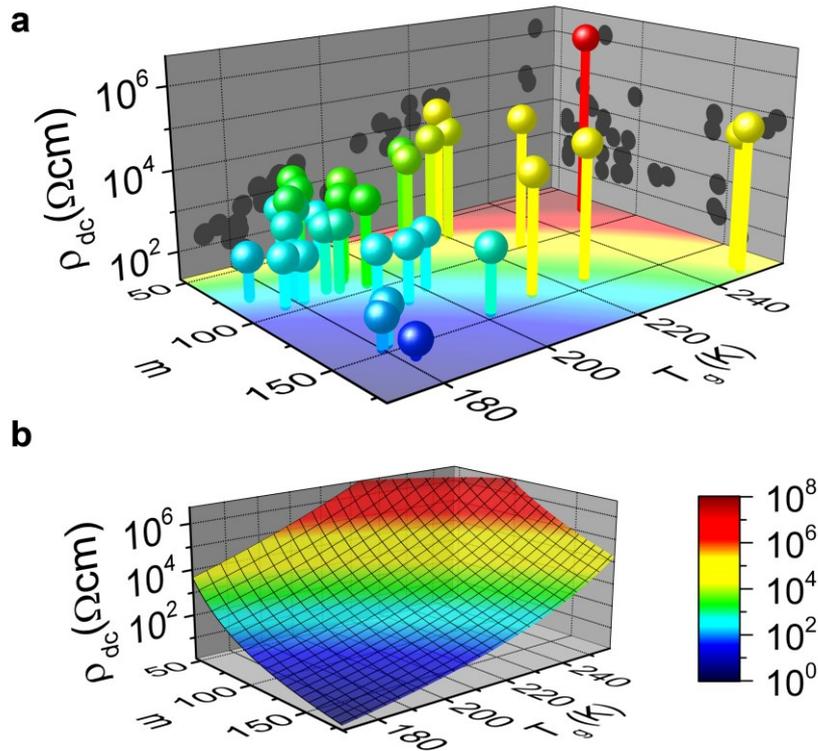

**Figure 4 | Correlation of the room-temperature dc resistivity of ionic liquids with the glass temperature and fragility.** In **a**, experimental data for the 13 ionic liquids measured by us and for another 22 compounds taken from literature are included (see Table 1). The spheres are colour coded as indicated in the colour bar. Frame **b** shows a colour-coded plane calculated from the VFT law making a number of assumptions as noted in the text. The colours shown in the $m$-$T_g$ plane of frame **a** correspond to the colours of the plane in **b**. The colours of the columns, connecting the data points in **a** to the $m$-$T_g$ plane, approximately match the colours of this plane. Thus, the calculation result shown in **b** roughly accounts for the experimentally observed absolute values of $\rho_{dc}$ in **a**.

In summary, the experimental data on 35 ILs, 13 of them investigated by dielectric spectroscopy in the present work, reveal a distinct dependence of the room-temperature resistivity on both glass temperature *and* fragility. We find that those ILs that combine high fragility with low glass temperature have the highest conductivity. Our work clearly demonstrates that fragility is just as essential for their optimisation for application as the glass temperature. Fragility is an old concept in glass physics. There are numerous approaches trying to explain what makes a glass former fragile, e.g., via energy-landscape variations or an increase of cooperativity[7,30]. For ILs, the dependences of fragility on ionicity[11] or anion size[16] have been considered. Following these lines seems a promising task on the way to the development of better ILs, suitable for electrochemical applications.

## Methods

**Sample preparation.** The samples were purchased from IoLiTec (Ionic Liquids Technologies GmbH, Heilbronn, Germany) with a minimum purity of 97%. To minimize water content, all samples were dried in $N_2$-gas or vacuum at elevated temperatures for several hours right before measurement. The sample of [1,3-Dimethylimidazolium + Li 1.0mol/kg] bis(trifluoromethane)sulfonamide was prepared by Dr. Xiao-Guang Sun from Oak Ridge National Laboratory and had a purity > 98%.

**Dielectric measurements.** Different experimental techniques were combined to measure the dielectric properties over a broad frequency range of up to ten decades (0.1 Hz $< \nu <$ 3 GHz). In the low-frequency range, $\nu <$ 1 MHz, a frequency-response analyser (Novocontrol Alpha-analyser) and an autobalance bridge (Agilent 4980A) where used. Measurements at $\nu >$ 1 MHz were performed by a I-V technique where the sample capacitor is mounted at the end of a coaxial line, bridging inner and outer conductor[26]. For these measurements, impedance analysers (Agilent E4991A or Hewlett-Packard HP4291A) were used. For both methods, the sample materials were filled into parallel-plate capacitors. For cooling and heating of the samples, a $N_2$-gas cryostat was used.

## Acknowledgements
We thank Dr. Changwoo Do and Dr. Xiao-Guang sun from Oak Ridge National Laboratory for kindly providing the sample of [1,3-Dimethylimidazolium + Li 1.0mol/kg] bis(trifluoromethane)sulfonamide. This work was supported by the Deutsche Forschungsgemeinschaft via Research Unit FOR1394 and by the BMBF via ENREKON.

## Author contributions
A.L., P.L. and S.K. conceived and supervised the project. P.S. and E.T. performed the dielectric measurements. P.S. analysed the data. P.L. wrote the paper. All authors discussed the results and commented on the manuscript. We thank M. Aumüller, L. Haupt and M. Weiss for performing parts of the dielectric measurements.